\documentclass[12pt,preprint]{aastex}

\begin{document}
\def\ale{\mathrel{\hbox{\rlap{\hbox{\lower4pt\hbox{$\sim$}}}\hbox{$<$}}}}
\def\age{\mathrel{\hbox{\rlap{\hbox{\lower4pt\hbox{$\sim$}}}\hbox{$>$}}}}

\title{Discovery of GRB~020405 and its Late Red Bump}

\author{
P.~A.~Price\altaffilmark{1,2},
S.~R.~Kulkarni\altaffilmark{2},
E.~Berger\altaffilmark{2},
D.~W.~Fox\altaffilmark{2},
J.~S.~Bloom\altaffilmark{2},
S.~G.~Djorgovski\altaffilmark{2},
D.~A.~Frail\altaffilmark{3},
T.~J.~Galama\altaffilmark{2},
F.~A.~Harrison\altaffilmark{2},
P.~McCarthy\altaffilmark{4},
D.~E.~Reichart\altaffilmark{2},
R.~Sari\altaffilmark{5},
S.~A.~Yost\altaffilmark{2},
H.~Jerjen\altaffilmark{1},
K.~Flint\altaffilmark{6},
A.~Phillips\altaffilmark{7},
B.~E.~Warren\altaffilmark{1},
T.~S.~Axelrod\altaffilmark{1},
R.~A.~Chevalier\altaffilmark{8},
J.~Holtzman\altaffilmark{9},
R.~A.~Kimble\altaffilmark{10},
B.~P.~Schmidt\altaffilmark{1},       
J.~C.~Wheeler\altaffilmark{11},
F.~Frontera\altaffilmark{12,13},
E.~Costa\altaffilmark{12},
L.~Piro\altaffilmark{12},
K.~Hurley\altaffilmark{14},         
T.~Cline\altaffilmark{15},
C.~Guidorzi\altaffilmark{13},         
E.~Montanari\altaffilmark{13},         
E.~Mazets\altaffilmark{16},
S.~Golenetskii\altaffilmark{16},
I.~Mitrofanov\altaffilmark{17},
D.~Anfimov\altaffilmark{17},
A.~Kozyrev\altaffilmark{17},
M.~Litvak\altaffilmark{17},
A.~Sanin\altaffilmark{17},
W.~Boynton\altaffilmark{18},
C.~Fellows\altaffilmark{18},
K.~Harshman\altaffilmark{18},
C.~Shinohara\altaffilmark{18},
A.~Gal-Yam\altaffilmark{19},
E.~Ofek\altaffilmark{19} and
Y.~Lipkin\altaffilmark{19}.
}

\altaffiltext{1}{Research School of Astronomy \& Astrophysics, Mount Stromlo
Observatory, via Cotter Road, Weston, ACT, 2611, Australia.}
\altaffiltext{2}{Palomar Observatory, 105-24, California Institute of
Technology, Pasadena, CA, 91125.}
\altaffiltext{3}{National Radio Astronomy Observatory, P.O. Box O, Socorro,
NM, 87801.}
\altaffiltext{4}{Carnegie Observatories, 813 Santa Barbara Street, Pasadena,
CA 91101.}
\altaffiltext{5}{Theoretical Astrophysics, 130-33, California Institute
of Technology, Pasadena, CA, 91125.}
\altaffiltext{6}{UCO/Lick Observatory, Board of Studies in Astronomy and
Astrophysics, University of California, Santa Cruz, 1156 High Street, Santa
Cruz, CA 95064.}
\altaffiltext{7}{School of Physics, The University of New South Wales,
Sydney, NSW 2052, Australia.}
\altaffiltext{8}{Department of Astronomy, University of Virginia,
P.O. Box 3818, Charlottesville, VA 22903-0818.}
\altaffiltext{9}{Department of Astronomy, MSC 4500, New Mexico State
University, P.O.~Box 30001, Las Cruces, NM 88003.}
\altaffiltext{10}{Laboratory for Astronomy and Solar Physics, NASA Goddard
Space Flight Center, Code 681, Greenbelt, MD 20771.}
\altaffiltext{11}{Astronomy Department, University of Texas, Austin, TX 78712.}
\altaffiltext{12}{Istituto Astrofisica Spaziale e Fisica Cosmica, C.N.R.,
Area di Tor Vergata, Via Fosso del Cavaliere 100, 00133 Roma, Italy.}
\altaffiltext{13}{Dipartimento di Fisica, Universita di Ferrara,
Via Paradiso 12, 44100, Ferrara, Italy.}
\altaffiltext{14}{University of California Space Sciences Laboratory,
Berkeley, CA, 94720.}
\altaffiltext{15}{NASA Goddard Space Flight Center, Code 661, Greenbelt,
MD 20771.}
\altaffiltext{16}{Ioffe Physico-Technical Institute, Saint Petersburg 194021,
Russia.}
\altaffiltext{17}{Space Research Institute, Profsojuznaya Str. 84/32, 117810,
Moscow, Russia.}
\altaffiltext{18}{Department of Planetary Sciences, Lunar and Planetary Laboratory, Tucson, AZ 85721-0092.}
\altaffiltext{19}{School of Physics \& Astronomy and Wise Observatory,
Tel-Aviv University, Tel-Aviv 69978, Israel.}


\begin{abstract}
We present the discovery of GRB~020405 made with the Inter-Planetary
Network (IPN).  With a duration of 60 s, the burst appears to be a
typical long duration event.  We observed the 75-square acrminute IPN
error region with the Mount Stromlo Observatory's 50-inch robotic
telescope and discovered a transient source which subsequently decayed
and was also associated with a variable radio source.  We identify
this source as the afterglow of GRB\,020405.  Subsequent observations
by other groups found varying polarized flux and established a
redshift of 0.690 to the host galaxy. Motivated by the low redshift we
triggered observations with WFPC2 on-board the {\it Hubble Space
Telescope} (HST).  Modeling the early ground-based data with a jet
model, we find a clear red excess over the decaying optical
lightcurves that is present between day 10 and day 141 (the last HST
epoch).  This ``bump'' has the spectral and temporal features expected
of an underlying supernova (SN).  In particular, the red color of the
putative SN is similar to that of the SN associated with GRB\,011121,
at late time.  Restricting the sample of GRBs to those with $z<0.7$, a
total of five bursts, red bumps at late times are found in
GRB\,970228, GRB\,011121, and GRB\,020405. It is possible that the
simplest idea, namely that all long duration GRBs have underlying SNe
with a modest dispersion in their properties (especially peak
luminosity), is sufficient to explain the non detections.
\end{abstract}



\section{Introduction}
\label{sec:intro}

In recent years several indirect lines of evidence have emerged
connecting the class of long duration $\gamma$-ray bursts (GRBs) to
massive stars.  Every GRB afterglow with a sub-arcsecond position is
associated with a star-forming galaxy \citep{bkd02}.  Some of these
galaxies are forming stars copiously with rates of a few hundred
M$_\odot$ yr$^{-1}$ \citep{bkf01,fbm+02}.  On smaller scales, some
afterglows (the so-called ``dark bursts''), show evidence for heavy
dust extinction \citep{dfk+01,pfg+02}.  X-ray and optical observations
of some GRBs indicate substantial column densities
\citep{ogo+98,gw01}.  In addition, there is evidence for moderate
circumburst densities, $n\sim 10$ cm$^{-3}$, in some bursts
\citep{hys+01,pk01,yfh+02}.  These indicators are consistent with GRBs
originating in gas-rich star-forming regions (i.e.~molecular clouds).

The most direct link between GRBs and massive stars comes from
observations on stellar scales, namely the detection of underlying
supernovae (SNe) and X-ray spectral features.  X-ray spectral features
have been observed in a few GRBs (e.g.~\citealt{pgg+00,rwo+02}),
although the detections generally have low signal-to-noise, and the
interpretations are somewhat controversial.  What is generally agreed,
however, is that X-ray features would require the presence of high
densities of iron on stellar scales.

The discovery of the unusual Type Ic SN~1998bw \citep{gvv+98} 
in a nearby ($\sim$40~Mpc) galaxy, within the small error box of 
GRB~980425 \citep{paa+00} suggested that at least some GRBs might be
caused by SN explosions.  Despite the fact that GRB~980425 was
under-energetic compared to the cosmological GRBs \citep{fks+01} and
may therefore represent an independent class of GRBs, the fact remains
that SN~1998bw directly demonstrates that a massive star is capable of
producing relativistic ejecta \citep{kfw+98} --- an essential
requirement for producing $\gamma$-rays.

The first indication of a SN underlying a cosmological GRB came from
the observation of a red excess (``bump'') in the rapidly-decaying
afterglow of GRB~980326 \citep{bkd+99}, which had a color and peak
time consistent with SN~1998bw shifted to $z\sim 1$.  However, the
lack of a measured redshift for this GRB and the possibility of other
explanations (e.g. dust echoes: \citealt{eb00,reichart01b}) made the
identification of the bump uncertain.  Several attempts to identify
similar bumps in the afterglows of GRBs with known redshift followed
with mixed results; see \citet{pks+02} for a review.

These earlier results motivated us to successfully propose a large
{\it Hubble Space Telescope} (HST) program to search for SNe
underlying GRBs (GO-9180, P.I.: Kulkarni).  HST is ideally suited to
this effort since its stable point-spread function and high angular
resolution make possible accurate and precise photometry of variable
sources embedded on host galaxies.  Low redshift GRBs are particularly
important to study, since beyond a redshift of 1.2 the strong
absorption in the SNe rest-frame spectra blueward of 4000 \AA\ covers
the entire observed optical region, thus making searches all but
impossible with current instruments.

To date, the best case for a SN underlying a cosmological GRB comes
from HST observations of GRB~011121 ($z=0.365$;
\citealt{bkp+02,gsw+02}).  This is based on a bump in the optical
afterglow lightcurves between 15 and 75 days, exhibiting a spectral
turnover at $\sim$ 7200 \AA.

In addition, based on early NIR and radio observations \citep{pbr+02},
the afterglow of GRB~011121 exhibits clear evidence for a circumburst
density, $\rho\propto r^{-2}$ (where $r$ is the radial distance from
the burst).  Such a density profile is indicative of stellar mass
loss.  Hence, from two independent lines of evidence, it can be
inferred that the progenitor of GRB~011121 was a massive star.  

However, not all GRBs have an underlying SN as bright as that of
GRB~011121 (e.g. GRB~010921: \citealt{pks+02}).  Thus, additional
deeper searches for coincident SNe are necessary to determine whether
the lack of an observed SN is due to dust obscuration, diversity in
the brightness of SNe coincident with GRBs, or to some subset of GRBs
having a different progenitor.

So far, our discussion has been motivated by, and based on
observations.  However, theorists have studied massive star models for
long duration GRBs for more than a decade.  In particular, the
``collapsar'' model posits that GRBs arise when the cores of massive
stars with sufficient angular momentum collapse and form black holes
\citep{woosley93,mw99,mwh01} whose accretion powers bursts of
$\gamma$-rays.  From the discussion in this section, it is clear that
there is a good observational basis for the collapsar model.  Detailed
studies of the underlying SNe (or their absence) will provide much
needed observational constraints to the collapsar model, or other
models which also require an associated SN event (``supranova'' ---
\citealt{vs98}; ``cannonball'' --- \citealt{ddd02}).

Here, we present the discovery of the afterglow of GRB~020405 and the
subsequent search for and discovery with HST of a red bump in the
afterglow that we suggest may be a SN underlying the GRB.

\section{The GRB and its Optical Afterglow}
\label{sec:obs}

\begin{figure}
\plotone{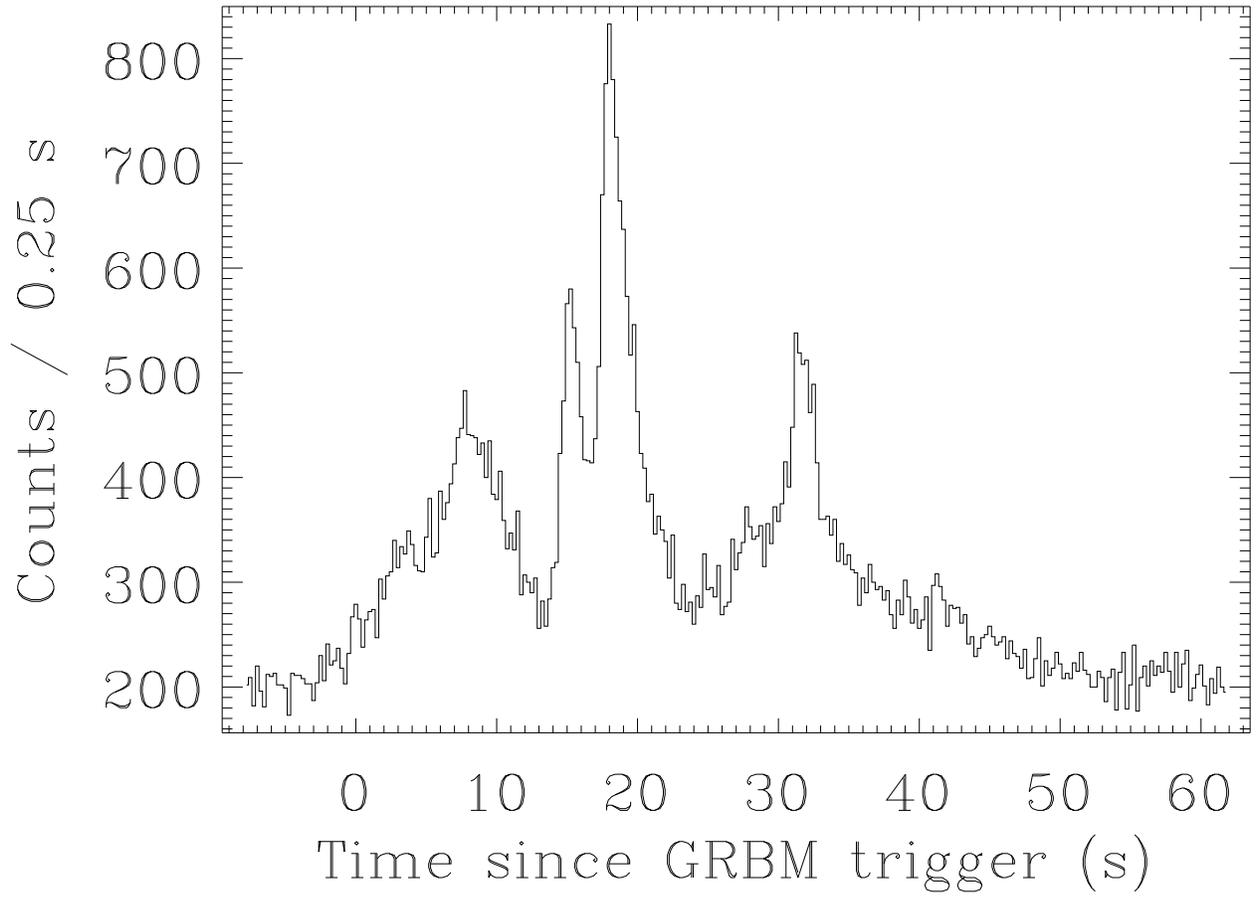}
\caption{The time history of GRB~020405, as observed by BeppoSAX
(40--700~keV).}
\label{fig:th}
\end{figure}

On 5 April 2002 at 00:41:26 UT the InterPlanetary Network (IPN)
consisting of Ulysses, Mars Odyssey/HEND and BeppoSAX discovered and
localized GRB~020405 \citep{hcf+02}.  With a duration of 60~s, the
burst is a typical long duration GRB (Figure~\ref{fig:th}).  The
prompt emission can be well fitted by a Band function \citep{bmf+93}
with the following parameters: low-energy spectral index,
$\alpha=-0.00\pm 0.25$, high-energy spectral index, $\beta=-1.87\pm
0.23$ and break energy, $E_b\sim 182\pm 45$ keV.  The fluence, as
measured by Konus-WIND, was $(7.40 \pm 0.07) \times 10^{-5}$
erg~cm$^{-2}$ (15--2000 keV), and the peak flux, averaged over
0.768~s, was $(5.0 \pm 0.2) \times 10^{-6}$ erg~cm$^{-2}$~s$^{-1}$.

We observed the 75 square-arcmin error box of GRB~020405 with the
robotic 50-inch telescope at Mount Stromlo Observatory (MSO) and the
40-inch telescope at Siding Spring Observatory (SSO), commencing
approximately 17 hours after the GRB.  From comparison of these images
with images from the Digitised Sky Survey\footnote{The Digitized Sky
Surveys were produced at the Space Telescope Science Institute under
U.S. Government grant NAG W-2166.}  we were able to identify a bright
($R \sim 18.5$ mag) source within the error-box that was not present
in the Sky Survey \citep{psa02}.  We undertook further imaging
observations with the Wise 40-inch, SSO 2.3-m, and the Las Campanas
du~Pont 100-inch telescopes (see Table~\ref{tab:ground}) and found
that the candidate faded.

In parallel, we undertook VLA observations of the source and found a
$0.4$ mJy radio counterpart \citep{bkf02}.  The combination of a
decaying optical source and a variable radio counterpart established
that we had detected the afterglow of GRB~020405.

Several groups undertook follow-up observations of the optical
afterglow.  In particular, observations with the VLT rapidly
identified the host galaxy to be a member of a group of interacting
galaxies \citep{mpm+02} at a redshift of $z = 0.695\pm 0.005$
\citep{mpp+02}.

\section{Optical Observations of the Afterglow and the Host Galaxy} 

The low redshift of this event made it a prime candidate for a search
for an underlying SN as a part of our large HST program.  We staggered
observations with the Wide Field Planetary Camera 2 (WFPC2) in F555W,
F702W and F814W between 19 and 31 days after the GRB in order to
densely sample the peak of any underlying SN.  We followed this
sequence with observations in each filter approximately two months
after the GRB in order to measure the SN decay.  At each epoch we
exposed a total of 3900~s and used a 6-point dither pattern to recover
the under-sampled point-spread function (PSF).

We used ``On-The-Fly'' pre-processing to produce debiased, flattened
images.  The images were then drizzled \citep{fh02} onto an image with
pixels smaller than the original by a factor of 0.5 using a {\tt
pixfrac} of 0.7.

Given that we had to tie the ground based and HST datasets we paid
special attention to calibration.  Calibration of field stars was
performed through observation of Stetson standards\footnote{\tt
http://cadcwww.dao.nrc.ca/standards/ } with the Swope 40-inch
telescope at LCO.  We estimate that our calibration is accurate to
approximately 0.05~mag.  Images were bias-subtracted and flat-fielded
in the standard manner, and combined where necessary to increase the
signal-to-noise.

\begin{figure}
\includegraphics[angle=-90,scale=0.75]{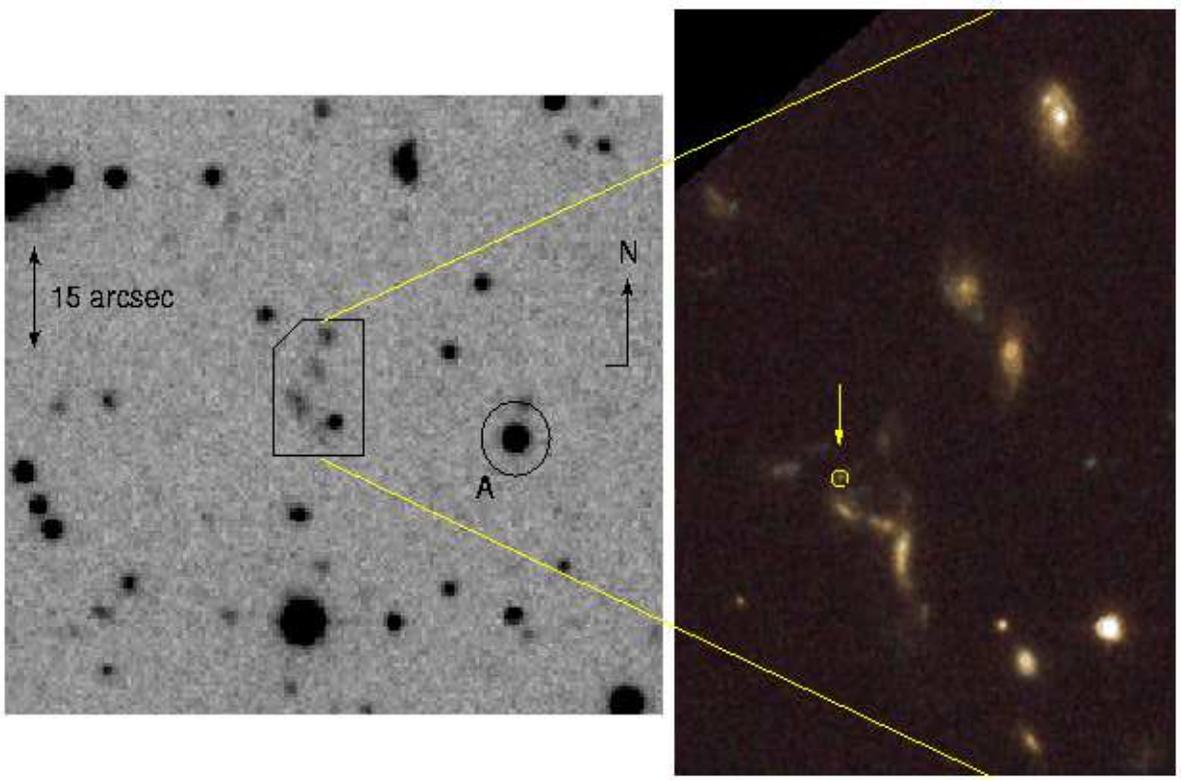}
\caption{SSO 2.3m (left) and HST (right) images of the host galaxy
complex of GRB~020405.  The GRB is 37.10'' East and 6.69'' North of
the star marked ``A'', for which we measure: $B = 19.787 \pm 0.017$
mag, $V = 18.945 \pm 0.016$ mag, $R = 18.452 \pm 0.008$ mag and $I =
17.980 \pm 0.010$ mag.  The position of the GRB in the HST image is
labeled.  The host complex is relatively bright ($R \sim 21$ mag).}
\label{fig:finder}
\end{figure}

As can be seen from Figure~\ref{fig:finder}, the host galaxy complex
is quite bright ($R \sim 21$ mag). Hence it is essential to employ
image subtraction techniques on the ground-based images to obtain
accurate lightcurves for the afterglow (e.g.
\citealt{pks+02,bmg+02}).  We used the Novicki-Tonry photometry
technique (see \citealt{pks+02} for a description) on both the
ground-based and HST observations to produce host-subtracted fluxes,
under the assumption that the afterglow flux in the last available HST
images in each filter were negligible (Tables~\ref{tab:ground} and
\ref{tab:hst}).  For the ground-based data, this is a reasonable
assumption, since none of the observations are particularly deep, and
the errors in the photometry will be larger than any offset.
Similarly for the F555W and F702W measurements with HST, which have
late observations, but in F814W there will be an uncertain, additional
flux not included in our measurements since we are using subtracted
fluxes.  This additional flux is equal to the flux of the afterglow
on Jun 9 in F814W, and so it is important to bear in mind that the
F814W measurements presented below are an underestimate of the true
flux for this filter only.

\subsection{Host Galaxy Spectroscopy}

We observed the afterglow and host galaxy of GRB~020405 with the
Echellette Spectrograph and Imager \citep{sbe+02} on Keck II at 2002
Apr~8.41~UT in poor seeing conditions.  We used a 0.75-arcsec wide
slit close to the parallactic angle and obtained two 1800-s exposures
in echellette mode with a small dither on the slit.  We used custom
software to straighten the echelle orders before combining the
individual exposures and extracting the spectrum using IRAF.  Arc lamp
exposures were used for wavelength calibration, with a resultant
scatter of 0.06 \AA.  An observation of Feige 34 was used for flux
calibration.  An earlier spectrum was also obtained from the Baade
Telescope, but with lower resolution and signal-to-noise.

We detect several bright emission lines which we attribute to [O~II],
H$\beta$ and [O~III] at a mean Heliocentric redshift of 0.68986 $\pm$
0.00004, consistent with the measurement first reported by
\citet{mpp+02}.  We list these emission lines in Table~\ref{tab:spec}.
Using the observed [O~II] and H$\beta$ line fluxes and assuming a flat
Lambda cosmology with $H_0$ = 65 km s$^{-1}$ Mpc$^{-1}$ and
$\Omega_{\rm M}$ = 0.3, we calculate \citep{ken98} a star formation
rate of 4 M$_\odot$ yr$^{-1}$.  This star-formation rate is
uncorrected for extinction or stellar absorption, and is therefore a
lower limit.

\section{Modeling the Afterglow}

\begin{figure}[tbp]
\plotone{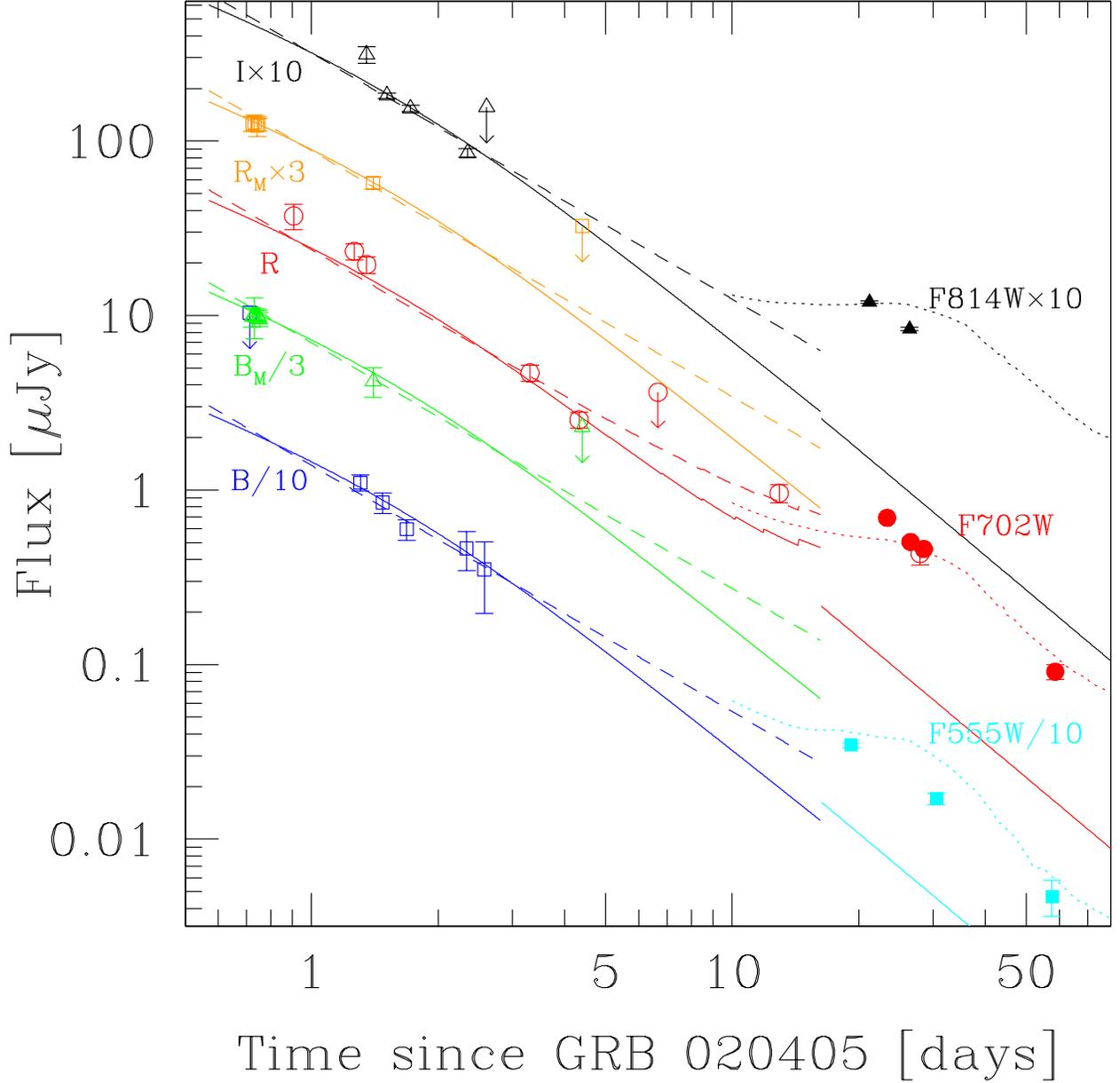}
\caption{Light curves of the optical afterglow of GRB~020405, assuming
zero afterglow flux in the final HST measurements.  Open points are
used for ground-based measurements, filled points for HST
measurements.  The dashed line is a single power-law decay model
(isotropic emission).  The solid line is a broken power-law decay
model (jet).  Both models incorporate a power-law spectrum and are fit
to data taken before 10 days.  We have plotted the light curve of
SN~1998bw shifted to $z=0.690$ and dimmed by 0.5~mag over the HST data
for a rough comparison.  The flux in the F814W filter is an
underestimate; see the text for an explanation.  Reddening the
SN~1998bw light-curve to account for host extinction may produce a
better match, but the extinction along the line-of-sight cannot be
precisely determined from the current data.}
\label{fig:lc}
\end{figure}

We model the lightcurves (Figure~\ref{fig:lc}) by adopting a standard
afterglow model with power-law temporal decay and a power-law
spectrum, $F_\nu \propto t^\alpha \nu^\beta$.  Due to the bright host
galaxy complex, we include in our data set only those values in the
literature that have been derived from host-subtracted images ---
specifically, the measurements of \citet{bmg+02} (corrected for the
apparent difference in reference star magnitudes) and those presented
here.  In the first round of analysis we restrict ourselves to data
taken prior to 10 d after the GRB.  We obtain a fairly poor fit with
$\chi^2$/DOF = 53.4/29, $\alpha = -1.41$ and $\beta = -1.25$.  We
notice that the model slightly over-predicts the flux for the earliest
data from the MSO 50-inch.  This may suggest the presence of a jet
break about one day after the GRB.

\citet{bmg+02} have detected strong polarization of the afterglow
emission
(9.9\%) at 1.3 d after the GRB which then appears to have declined to
about 1.9\% by 2.1 d after the GRB \citep{cgm+02}.  \citet{s99}
predicted the polarization of GRB afterglows to peak at about 10\% at
the time of the jet break for GRBs that are not viewed far off-axis.
The behavior of the polarization-curve, therefore, also argues for
the presence of a jet break about a day after the GRB.  The low radio
flux (Berger et al.~in prep.) compared to the bright optical
afterglow may also be evidence for an early jet break.

We therefore adopt a broken power-law temporal decay with indices
$\alpha_1$ (early times), $\alpha_2$ (late times) and a jet break
time, $t_{\rm jet}$.  The power-law indices are functions of the
electron energy distribution index, $p$ ($N(\gamma)\propto
\gamma^{-p}$ for $\gamma>\gamma_{\rm min}$).  However, given the
sparse data we make the {\it ad hoc} simplifying assumption that the
optical band is above the cooling frequency and the circumburst medium
is homogenous.  In this case, $\alpha_1= (3p-2)/4$ and $\alpha_2 = -p$
\citep{sph99}.

With these assumptions, we obtain a much-improved fit ($\chi^2$/DOF =
35.9/28): $p = 1.93 \pm 0.25$ and $t_j = 1.67\pm 0.52$ d.  Adding a
systematic error of 0.06~mag in quadrature with the measurement errors
to account for differences between data taken with different
instruments reduces the $\chi^2$ to match the degrees of freedom.  In
our experience, this is an expected level of systematic error.

The redshift of GRB~020405, $z=0.690$, and the observed spectrum of
the burst, imply an isotropic-equivalent $k$-corrected \citep{bfs01}
energy release, $E_{\rm iso}(\gamma)[20-2000 {\rm keV}] = (7.37 \pm
0.80) \times 10^{52}$ erg.  The $k$-correction ($k=1.31 \pm 0.09$) is
small and rather precise given the spectral constraints from BeppoSAX.
From our best-fit value of $t_{\rm jet}$, using the specific
formulation of \citet{fks+01}, we calculate a jet opening angle of
$(5.83\pm 0.69)\,n^{1/8}$ degrees, where $n$ is the number density of
the ambient medium in units of 1 cm$^{-3}$.  The beaming-corrected
energy is thus $E_\gamma = (3.82 \pm 0.94)\times 10^{50}\,n^{1/4}$
erg, at the low end of (but consistent with) the distribution centred
on $9\times 10^{50}\,n^{1/4}$ erg for long duration GRBs
\citep{fks+01}.

We can also measure the dust extinction from the afterglow data by
demanding that the intrinsic spectral slope match that predicted by a
particular theoretical model (dependent upon the density profile and
the location of the cooling frequency relative to the optical bands)
and attributing any observed reddening to extinction (see, e.g.,
\citealt{pbr+02}).  Assuming an SMC extinction curve
\citep{reichart01a}, we measure the extinction to be 0.22 mag $<
A_V^{\rm host} <$ 0.64 mag, depending on the choice of afterglow
model.  

\subsection{Late Time Data: A Red Bump}

We now turn our attention to data taken after 10 d from the GRB.  We
note that there is a strong excess in each of the three HST filters.
This excess (``bump'') exists independent of the assumed geometry of
the afterglow, but is more pronounced for a jet model (discussed
above).  We suggest this excess may be due to a SN which exploded
within about a week of the GRB.  Although it is not possible to
analyze the bump in detail (in particular, its flux relative to
SN~1998bw) until later HST observations have been made of the host
galaxy to remove the assumption of zero afterglow flux in the last
available epochs, we can make a number of qualitative statements.

First, the peak of the bump is not well constrained by these data, and
appears to be between 10 and 25 days after the GRB.
Second, fitting a power-law spectrum, $F_\nu \propto \nu^\beta$, to
the HST data demonstrates that the bump is quite red, with $\beta =
-3.98 \pm 0.18$ (i.e.\ $B-V=1.07$~mag, $V-R=0.90$~mag), in contrast to
$\beta = -1.23 \pm 0.12$ measured for the afterglow at early times,
further evidence for the existence of two components.  This
measurement of the spectral index is similar to that for the SN
underlying GRB~011121, which has $\beta \approx -3.5$ between the
F450W and F555W filters at late time \citep{bkp+02}.  However, the SN
underlying GRB~011121 appeared somewhat bluer in the F555W filter than
SN~1998bw \citep{gsw+02}, while GRB~020405's red bump appears more
red.  For GRB~011121, the SN spectral peak was at $\sim 7500$ \AA\
\citep{bkp+02}, which corresponds to 9300~\AA\ at the redshift of
GRB~020405.  Since the spectrum is well described by a single
power-law, the spectral peak of a SN would be redward of 8700~\AA,
which is consistent with the data on GRB~011121.  Thus, the broad-band
spectra of the red bump appears to be grossly similar to that of
GRB~011121 upon first inspection.  Some differences in color are not
unexpected, due to the uncertain extinction along the line-of-sight to
GRB~020405 and the expected diversity in the properties (e.g.~jets) of
these energetic SNe.

\section{Conclusions}

Here we report the discovery of the nearby ($z\sim 0.7$) GRB~020405
and the subsequent discovery of the afterglow.  The GRB itself, with a
duration of 60~s, appears to be a typical long duration burst. The
optical afterglow data, spanning about 10 days, can be fitted with a
standard broken power law with a break time of about 1.5
d. Identifying this break with a jet we obtain a beaming-corrected
energy release of about $2\times 10^{50}\,$erg, typical of that
inferred for long duration GRBs.

Motivated by the low redshift, we undertook multi-color observations
of the afterglow with HST. We found an excess over the flux predicted 
by the modeling of the afterglow from ground based data.  The overall
broad-band spectrum of the bump as well as its temporal evolution
are most simply explained as due to an underlying SN which exploded 
at about the same time as the GRB.

In \citet{pks+02}, we summarize the searches for underlying SNe in
$z<1.2$ GRBs (the redshift restriction arising from the fact that the
searches are conducted in the optical band; see \S\ref{sec:intro}).
Including GRB~020405, there are 13 GRBs with $z<1.2$.  A strong case
for an underlying SN can be made for GRB~011121 ($z=0.365$;
\citealt{bkp+02,gsw+02}) and GRB~020405 ($z=0.690$).  A good case can
be made for GRB~970228 ($z=0.695$; \citealt{reichart99,gtv+00}) and
GRB~980326 ($z$ unknown; \citealt{bkd+99}).  At first blush this
appears to be a low yield and suggestive that either there is
diversity in the progenitors of long duration GRBs or in the
properties of underlying SNe, or both.

However, it is important to bear in mind that one of the two unique
signatures for a SN is the spectral rollover at short wavelengths,
namely below about 4000 \AA\ (see \citealt{bkd+99}).  Thus for $z\sim
1$, observations in the $R$ and $I$ bands are critical, whereas at
lower redshifts observations in $V$ and $R$ bands are critical.
The $I$ band is quite noisy for ground-based observations whereas most
afterglows are well observed in $R$ and $V$ bands.

Restricting to $z<0.7$ we find five GRBs (970228, 011121, 020405,
990712 and 010921), of which underlying SNe have been identified in
the first three, and possibly in GRB~990712 \citep{bhj+01}.  The limit
for an underlying SN in GRB~010921 is not very stringent
\citep{pks+02}; in particular, an underlying SN fainter by more than
2~mag relative to that of SN 1998bw (about as bright as typical SNe
Ib/c) would have escaped identification. It is thus premature to
conclude that we need several progenitors to cause GRBs. What we can
conclude, though, is that dense sampling in several bands of nearby
GRBs is likely to remain a productive activity.

\acknowledgements

BPS and PAP thank the ARC for supporting Australian GRB research.  GRB
research at Caltech (SRK, SGD, FAH, RS) is supported by grants from
NSF and NASA.  KH is grateful for Ulysses and IPN support under JPL
Contract 958056 and NASA Grant NAG5-11451.  Support for Proposal
HST-GO-09180.01-A was provided by NASA through a grant from the Space
Telescope Science Institute, which is operated by the Association of
Universities for Research in Astronomy, Inc., under NASA contract
NAS5-26555.  We thank M.~Pettini, N.~Reddy and C.~Steidel for
undertaking observations at Keck under the auspices of the Caltech ToO
program.  Finally, we thank the staff of MSO, SSO, Wise, LCO, and
Keck, and the STScI for their assistance.

\bibliographystyle{apj1b}
\bibliography{journals_apj,ref_grb020405}

\clearpage

\begin{deluxetable}{cccccc}
\footnotesize
\tablecolumns{6}
\tablewidth{0pt}
\tablecaption{\label{tab:spec}Lines identified in the spectrum of
the host galaxy of GRB~020405.}
\tablehead{\colhead{ $\lambda_{\rm obs}$ (\AA)} & \colhead{Line} & \colhead{$F_{\rm obs}$} & \colhead{EW (\AA)} & \colhead{GW (\AA)} & }
\startdata
6298.67 $\pm$ 0.10    &	[O II] 	& 55.7 $\pm$ 4.7 &	28.0 $\pm$ 4.7	& 2.08 $\pm$ 0.22	\\
6303.59 $\pm$ 0.04    &	[O II]	& 64.5 $\pm$ 3.7 &	31.8 $\pm$ 4.6	& 1.28 $\pm$ 0.09	\\
6539.7                &	[Ne III]	& 13.1 $\pm$ 5.1 &	7.0 $\pm$ 3.0	& 2.2	\\
7337.13 $\pm$ 0.24    &	H$\gamma$& 19.9 $\pm$ 5.3 &	11.6 $\pm$ 3.9	& 1.88 $\pm$ 0.60	\\
8217.42 $\pm$ 0.11    &	H$\beta$& 66.3 $\pm$ 5.7 &	54 $\pm$ 11	& 2.54 $\pm$ 0.24	\\
8382.25 $\pm$ 0.08    &	[O III]	& 51.1 $\pm$ 3.9 &	26.3 $\pm$ 5.3	& 2.29 $\pm$ 0.17	\\
8463.42 $\pm$ 0.02    &	[O III]	& 201.9 $\pm$ 3.7 &	71.7 $\pm$ 5.9	& 2.545 $\pm$ 0.052	\\
\enddata
\tablecomments{Left to right, the columns are the observed wavelength of
the line, line identification, observed flux corrected for Galactic
extinction using $E_{(B-V)} = 0.054$ in units of $10^{-18}$ erg/cm$^2$/s,
observed equivalent width (uncorrected for contamination by the afterglow),
and observed Gaussian width.  The line at $\lambda6303.59$ is affected by
a bright night sky line.  Given the suggestion that GRB host galaxies may
exhibit strong [Ne~III] emission \citep{bdk02} we have included an entry
for this line.}
\end{deluxetable}

\clearpage

\begin{deluxetable}{lcrc}
\footnotesize
\tablecolumns{5}
\tablewidth{0pt}
\tablecaption{Ground-based observations of the afterglow of GRB~020405.}
\tablehead{\colhead{Date (2002 UT)} & \colhead{Filter} & \colhead{Flux ($\mu$Jy)} & \colhead{Telescope}}
\startdata
Apr 5.742	& $B$	& 36 $\pm$ 34		& SSO40	\\
Apr 6.506	& $B$	& 8.5 $\pm$ 1.1		& SSO40	\\
Apr 6.713 	& $B$	& 5.97 $\pm$ 0.80	& SSO40	\\
Apr 7.367	& $B$	& 4.6 $\pm$ 1.2		& dP	\\
Apr 7.605 	& $B$	& 3.5 $\pm$ 1.6		& SSO40	\\
Apr 5.752 	& $B_M$	& 29.5 $\pm$ 3.7	& MSO50	\\
Apr 5.761 	& $B_M$	& 29.9 $\pm$ 7.9	& MSO50	\\
Apr 5.772 	& $B_M$	& 30.1 $\pm$ 2.5	& MSO50	\\
Apr 5.777 	& $B_M$	& 29.6 $\pm$ 2.5	& MSO50	\\
Apr 5.781 	& $B_M$	& 28.5 $\pm$ 2.6	& MSO50	\\
Apr 6.430 	& $B_M$	& 12.6 $\pm$ 2.4	& MSO50	\\
Apr 9.434 	& $B_M$	& 3.0 $\pm$ 1.9 	& MSO50	\\
Apr 5.752 	& $R_M$	& 41.9 $\pm$ 4.2	& MSO50	\\
Apr 5.761 	& $R_M$	& 42.4 $\pm$ 4.4	& MSO50	\\
Apr 5.772 	& $R_M$	& 40.8 $\pm$ 5.4	& MSO50	\\
Apr 5.777 	& $R_M$	& 41.5 $\pm$ 3.5	& MSO50	\\
Apr 6.430 	& $R_M$	& 19.1 $\pm$ 1.5	& MSO50	\\
Apr 9.434 	& $R_M$	& 5.3 $\pm$ 2.8 	& MSO50	\\
Apr 5.936	& $R$	& 37.2 $\pm$ 6.2 	& Wise40 \\
Apr 11.694 	& $R$	& 1.5 $\pm$ 1.1		& SSO23	\\
Apr 6.541 	& $I$	& 18.20 $\pm$ 0.64	& SSO40	\\
Apr 6.750  	& $I$	& 15.39 $\pm$ 0.63	& SSO40	\\
Apr 7.378	& $I$	& 8.53 $\pm$ 0.50	& dP	\\
Apr 7.639 	& $I$	& 4.9 $\pm$ 5.4 	& SSO40	\\
\enddata
\tablecomments{These data have been obtained by the Novicki-Tonry photometry
technique, and hence contain no contribution from the host galaxy.  Zero
points were set through photometry of several calibrated field stars
(sequence available upon request from {\tt pap@mso.anu.edu.au}).
These data have not been corrected for Galactic extinction.  Telescopes
are: SSO40 --- Siding Spring Observatory 40-inch; MSO50 --- Mount Stromlo
Observatory 50-inch (robotic, using MACHO filters; \citealt{bg99});
dP --- du Pont 100-inch telescope at Las Campanas Observatory;
Wise40 --- Wise Observatory 40-inch telescope; SSO23 --- Siding Spring
Observatory 2.3-m.}
\label{tab:ground}
\end{deluxetable}

\clearpage

\begin{deluxetable}{lcc}
\footnotesize
\tablecolumns{5}
\tablewidth{0pt}
\tablecaption{HST observations of the afterglow of GRB~020405.}
\tablehead{\colhead{Date (2002 UT)} & \colhead{Filter} & \colhead{Flux ($\mu$Jy)} }
\startdata
Apr 24.225	& F555W &	0.345 $\pm$ 0.011	\\
May 5.585	& F555W &	0.170 $\pm$ 0.013	\\
Jun 2.636	& F555W &	0.047 $\pm$ 0.011	\\
Aug 23.171	& F555W	&	\ldots			\\
Apr 28.388	& F702W &	0.693 $\pm$ 0.010	\\
May 1.574	& F702W &	0.504 $\pm$ 0.010	\\
May 3.579	& F702W &	0.459 $\pm$ 0.009	\\
Jun 1.568	& F702W &	0.091 $\pm$ 0.009	\\
Aug 23.374	& F702W	&	\ldots			\\
Apr 26.229	& F814W &	1.194 $\pm$ 0.020	\\
May 1.440	& F814W &	0.837 $\pm$ 0.020	\\
Jun 9.513	& F814W &	\ldots			\\
\enddata
\tablecomments{These data were obtained through Novicki-Tonry photometry
and hence contain no contribution from the host galaxy.  Those measurements
for which no flux is recorded are the last available images in each filter;
the flux of the afterglow is assumed to be zero in each of these for the
purposes of Novicki-Tonry photometry.  Counts were converted to fluxes
by using IRAF/{\tt synphot} to calculate the response to a constant flux
of 1 mJy; the resultant fluxes are hence analogous to AB magnitudes.
These measurements have been corrected for charge transfer (in-)efficiency
(CTE) using the prescription of \citet{dolphin00}.
}
\label{tab:hst}
\end{deluxetable}

\clearpage

\clearpage

\clearpage

\clearpage

\end{document}